# Economic and On Demand Brain Activity Analysis on Global Grids

R. Buyya[1], S. Date[2], Y. Mizuno-Matsumoto[3], S. Venugopal[1], and D. Abramson[4]

**Abstract**

The lack of computational power within an organization for analyzing scientific data, and the distribution of knowledge (by scientists) and technologies (advanced scientific devices) are two major problems commonly observed in scientific disciplines. One such scientific discipline is brain science. The analysis of brain activity data gathered from the MEG (Magnetoencephalography) instrument is an important research topic in medical science since it helps doctors in identifying symptoms of diseases. The data needs to be analyzed exhaustively to efficiently diagnose and analyze brain functions and requires access to large-scale computational resources. The potential platform for solving such resource intensive applications is the Grid. This paper describes a MEG data analysis system developed by us, leveraging Grid technologies, primarily Nimrod-G, Gridbus, and Globus. This paper explains the application of economy-based grid scheduling algorithms to the problem domain for on-demand processing of analysis jobs.

## 1. Introduction

The emergence of high speed networks has made it possible to share geographically distributed resources such as supercomputers, storage systems, databases and scientific instruments in order to gather, process and transfer data smoothly across different administrative domains. Aggregations of such distributed resources, called computational grids[1], provide computing power that has made it possible to solve large scale problems in science, engineering and commerce. Biological sciences have several computational and data intensive problems which have not been tackled satisfactorily for the want of adequate computing and storage resources. The analysis of data gathered by monitoring activity in the brain is one such application.

**Brain Activity Analysis**

Brain activity is measured by the Magnetoencephalography (MEG) instrument which measures the magnetic fields generated by the electrical activity in the brain. This method is more accurate than others such as Electroencephalography (EEG) and Electrocorticography (ECoG)[2]. Another advantage of MEG is that it is non-invasive.

The MEG instrument consists of a number of sensors which record information about brain activity. Currently, MEG helmets with over 200 sensors are already used to detect magnetic brain fields by means of a sensitive transducer technology called Superconducting Quantum Interference Device (SQUID). This provides a doctor with a host of data offering the finest temporal and the highest spatial resolution. Specialists can detect a disorder by observing the complex brain wave form and analysing the frequency content. The doctor has to separate the MEG data into signal classes. Each of these contains a certain frequency band which allows the localisation of the signal's source. To this end, wavelet cross-correlation analysis, developed at the Osaka University by Dr. Mizuno-Matsumoto et al. [3], is used. Unlike the traditional Fourier-based analysis, wavelet-based analysis has the capability to explore the frequency content without losing the time information of the original brain data. This process has to be performed for each pair of sensors. This analysis lacks the time information of the original brain data but provides the similarity between a pair of brain data every frequency spectrum. By focusing on a spectrum, one could track a signal that is within the spectrum. This enables mapping of brain activity to patient behaviour and is useful to diagnose several diseases.

[1] Grid Computing and Distributed Systems (GRIDS) Lab, Dept. of Computer Science and Software Engineering, University of Melbourne, Australia.

[2]Graduate School of Information Science and Technology, Department of Bioinformatic Engineering, Osaka University, Japan

[3]Department of Information Systems Engineering, Graduate School of Osaka University, Japan.

[4] School of Computer Science and Software Engineering, Monash University, Australia

However, there are certain problems with MEG. Due to the high cost of equipment and the required shielding against stray magnetic fields, there are only limited numbers of MEG instruments around the world. Plus, the instrument generates a huge amount of data, all of which are not analysed due to of lack of computing resources. For example, a 64-sensor MEG instrument would produce 0.9GB of data over a period of an hour. Such a task generates 7,257,600 analysis jobs and would take 102 days on a commodity computer with a PentiumIII/500MHz processor and 256MB of memory. Proper recovery of the patient depends on the results being available as soon as possible so that the doctor can start the drug regimen in the shortest possible time. While it is possible to use a supercomputer to analyse the brain data, every medical facility cannot be expected to have access to such a machine. Also, access to a much larger pool of resources and consequently, larger amounts of computational power is possible if the application can be deployed on the Grid. Furthermore, there is also the advantage that the medical data can be shared easily among the partnering doctors.

The rest of the paper is organized as follows: Section 2 describes the flow diagram for the analysis. Section 3 describes the architecture for the execution. Section 4 explains the process of grid-enabling this application. The scheduling experiments and the results of those are detailed in Section 5. Section 6 presents related work and the final section summarizes the paper along with suggestions for future works.

## 2. NeuroGrid Analysis Model

The NeuroGrid project aims to convert the existing brain activity analysis application into a parameter sweep application for executing jobs which perform wavelet cross-correlation analysis for each pair of sensors in parallel on distributed resources. The jobs are both computationally and data intensive. Also they are independent of each other thus making them perfect for being executed on the Grid. The flow diagram for distributed analysis of brain activity is shown in Fig. 1.

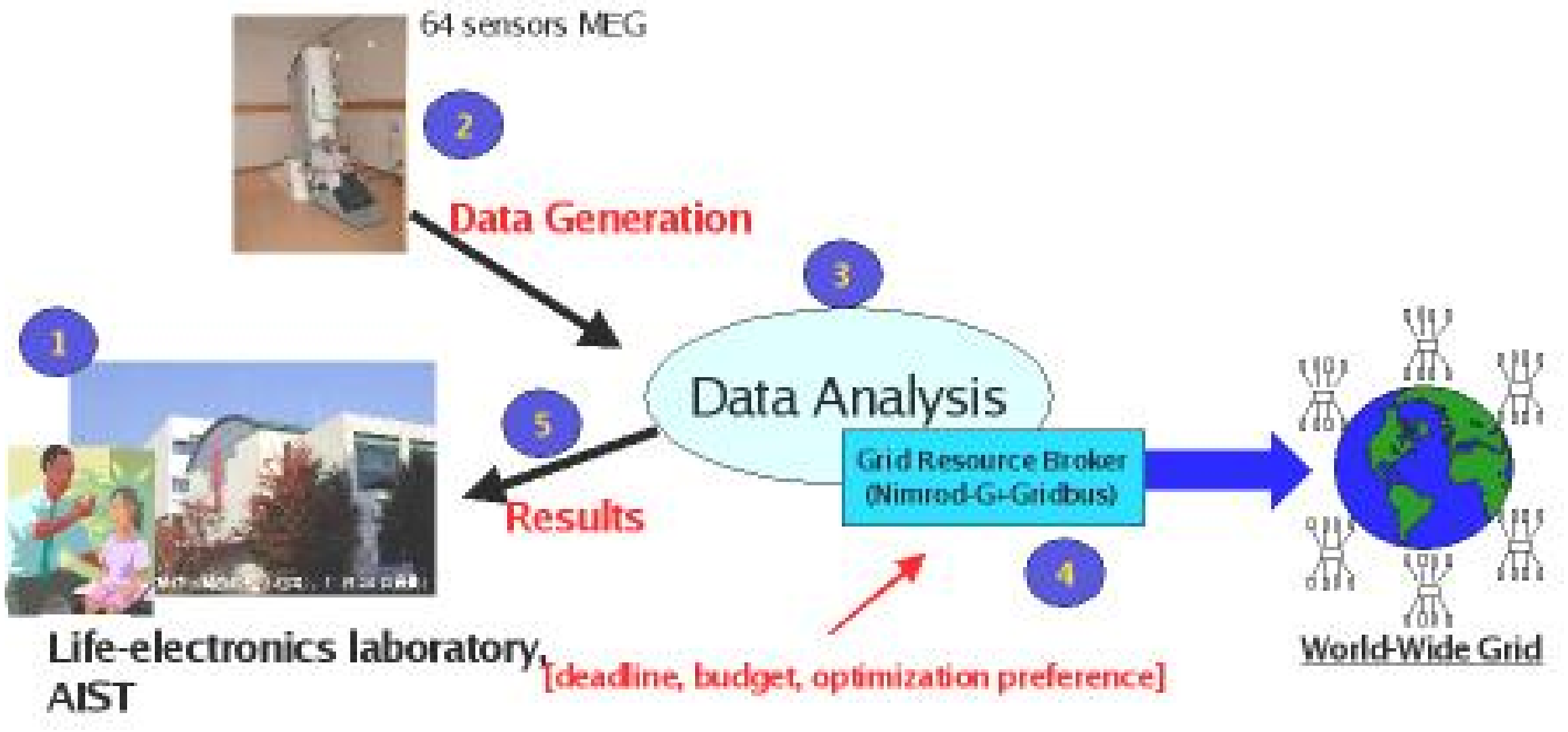

**Fig. 1: The Neuro Grid Analysis Model.**

The medical staff who is dealing with the diagnosis orders a MEG scan of the patient’s brain (step 1). The request is sent to the instrument which takes a MEG scan and collects data about the activity in the brain (step 2). This data is then collated and presented to the Grid Resource Broker for analyzing on the Grid (step 3).The broker discovers the resources and looks up in the Grid Market Directory for the corresponding service costs associated with those resources. Using the user-defined QoS (Quality-of-Service) parameters, two of which are the deadline and the budget, the scheduler distributes the jobs based on the optimization method chosen. The optimization method could be one of the three: cost, time or cost-time. The data and the analysis code are dispatched to the remote node and the results collected (steps 4 and 5).

## 3. Architecture

The architecture followed in this project is shown in Fig. 2. It consists of brain activity analysis application, paramterisation tools (Nimrod-G parameter specification language), resource broker (Nimrod-G with Gridbus scheduler), grid market directory (Gridbus GMD), and low-level grid middleware (Globus[4]). The resources are Grid-enabled using Globus software deployed on them. The application has been Grid enabled by composing it as a parameter sweep application using the Nimrod-G parameter specification language. The GMD has been used as a register for publication of resource providers and services.

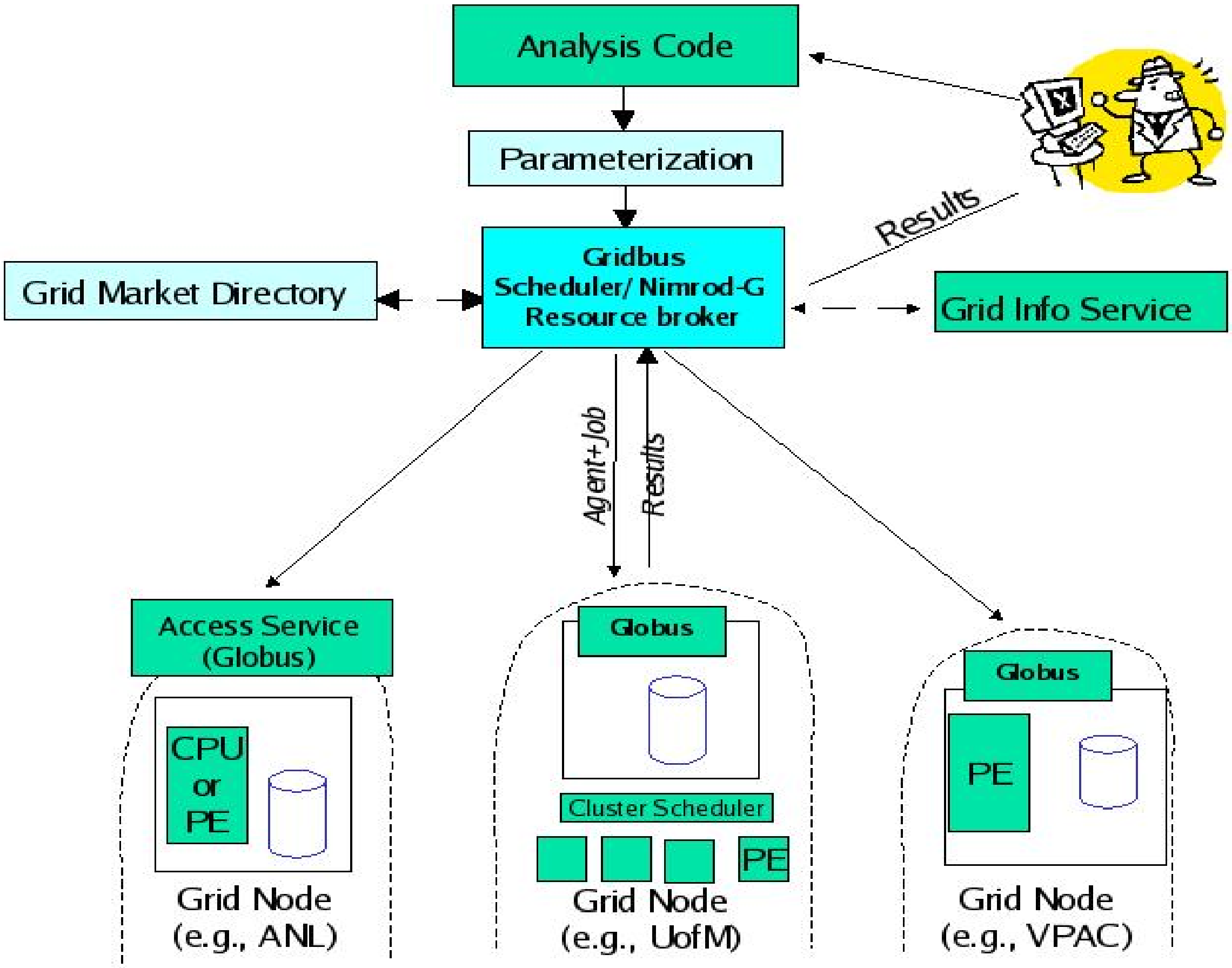

**Fig. 2: Architecture for Brain Activity Analysis on the Grid.**

### 3.1 Analysis Code

The analysis code was developed by the Cybermedia Centre, Osaka University, Japan. The raw data obtained from the sensors in the MEG instrument is analysed in two phases as shown in the Fig. 3. In the first phase, the raw data from the brain goes through wavelet transform operation. This phase gives the time-frequency data of the output. In the next phase, cross-correlation analysis is performed for each pair of wavelet transforms. This output displays the similarity between a pair of brain data in every frequency spectrum. By focusing on a specific frequency spectrum, the sensor that first detected the signal with that focusing spectrum can be localized. This means that the path of a signal can be found as it travels through the brain. The code was written in C and is highly portable. We have been able to compile it for different platforms such as PC/Linux, Sun/Solaris, SGI/Irix, Alpha, etc.

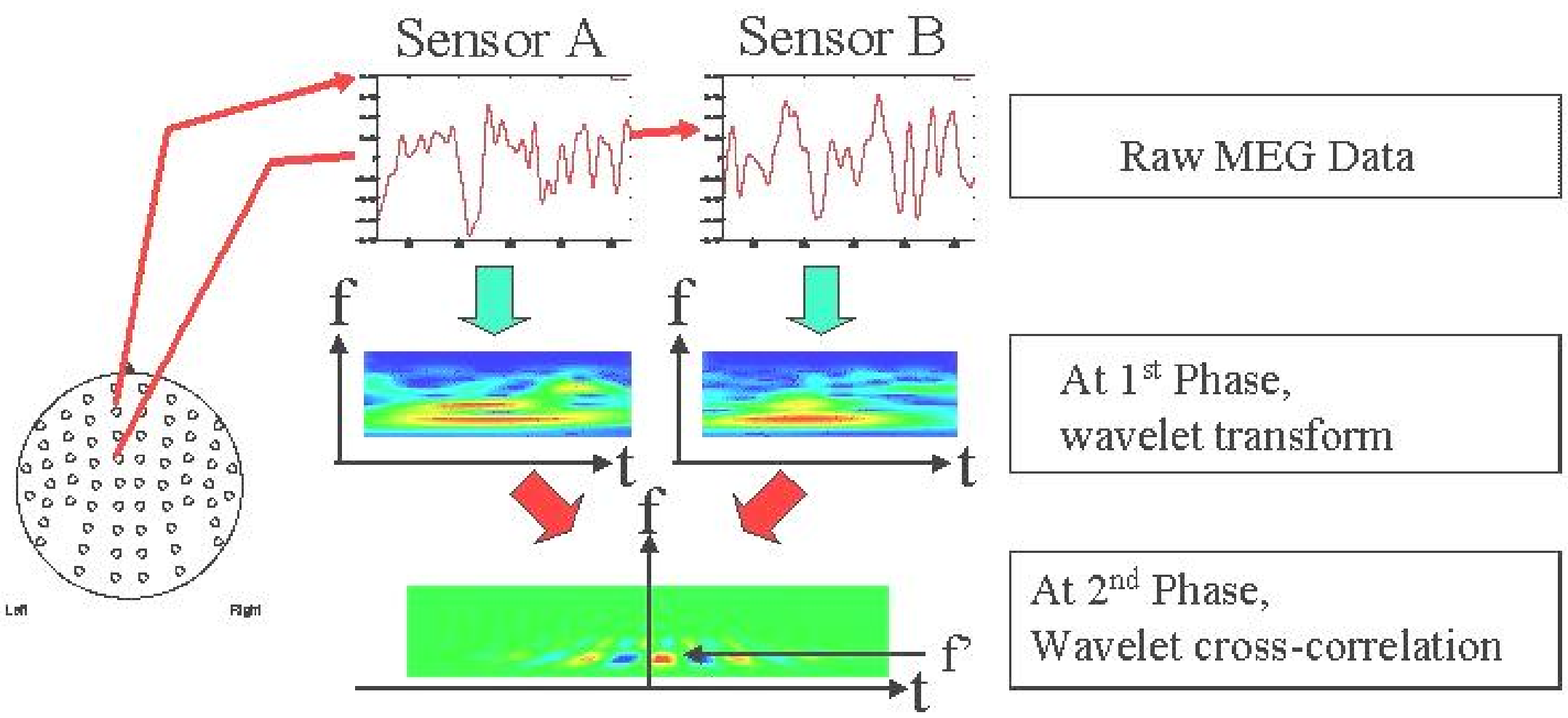

**Fig. 3: Wavelet cross-correlational analysis.**

### 3.2 Grid Resource Broker and Scheduler

For executing our application on the Grid, a combination of the Nimrod-G [5] resource broker developed using the Globus [4] middleware and our own Gridbus scheduler was used. The Nimrod-G resource broker identifies the user and application processing requirements and selects the combination of Grid resources in such a way that the user requirements are met. It performs resource discovery, selection, and dispatching of MEG jobs to remote resources. It also starts and manages the execution of jobs and gathers the results back at the home node. The following components of Nimrod-G have been used in our experiments:

- a persistent task farming engine;
- a grid explorer for resource discovery;

The Gridbus scheduler developed as a plugin scheduler for Nimrod-G has been used instead of the default Nimrod-G scheduler since it has been designed to utilise the Grid Market Directory (GMD) [6].The default Nimrod-G scheduler calculates the processing cost based on the CPU time that is used to execute a job on a remote node. To support the notion of application services and pricing based on AO (Application Operation) instead of vanilla CPU service, the GMD already allows the GSPs (Grid Service Providers) to publish application services along with their AO service price. Hence, the Gridbus scheduler has been enabled to utilise the GMD services and perform resource allocation based on AO cost model. In this model, the user is charged a price for execution of each job on the resource. Thus, the resource owner may offer the application as a service and charge a fixed price for executing it. By periodically updating the resource information from the GMD, the Gridbus scheduler ensures that the scheduling is done based on the most recent cost-price for the resources.

The Gridbus Scheduler implements three algorithms: cost minimization, time minimization and cost-time optimization. All three algorithms are constrained by two parameters: the deadline by which the experiment is required to complete and the budget that the user has. Time minimization tries to execute the project within the shortest time while keeping within the budget. Cost minimization tries to complete the execution with the least cost while keeping to the deadline. Cost-time optimization gives jobs to the cheapest servers but performs time optimization among those. Cost-time optimization was only simulated until recently and it has been implemented for the first time in the Gridbus Scheduler. More information about these scheduling algorithms can be found in [7, 8]. The scheduler uses the past performance of each machine to forecast the job completion rate of that machine. Also, it averages the job completion rate so that any spikes or troughs in performance are smoothed out.

### 3.3 Grid Market Directory

The Grid Market Directory (GMD)[6], developed by the Gridbus project in the University of Melbourne, allows service providers to publish the services which they provide along with the costs associated with them which the consumers are able to browse and locate a service which meets their requirements. GMD is built over standard Web service technologies such as SOAP (Simple Object Access Protocol) and XML(eXtensible Markup Language) therefore it can be queried by other client programs. To provide with an additional layer of transparency, a client API (Application Programming Interface) has been provided that could be used by programs to query the GMD without the developers having to concern themselves with details of SOAP. The GMD infrastructure was used to maintain a registry of participants along with the details of their contributed resources as part of the Global Grid Testbed collaboration. The Gridbus scheduler interacts with the GMD to discover the testbed resources and their high-level attributes such as access price.

## 4. Grid-enabling the application

The existing analysis code was composed as a task-farming, parameter sweep application for execution on the Grid using the Nimrod-G parameter specification language[5]. The Nimrod-G farming engine and dispatcher along with Gridbus scheduler is used for deploying and processing it on Global Grids.

The brain activity analysis software suite, developed at the Cybermedia Centre at Osaka University, consists of two separate programs. The first, *raw2wavelet*, is a program that performs wavelet transform over the raw data and the second, *wavelet2cross*, is a program that does the cross-correlation between the wavelet transforms. A job consists of executing these two programs in sequence for a pair of unique sensors and for a particular offset. For 64 sensors and an offset ranging over 0 to 29,750, this would generate (64*63)/2 * 29750 jobs. Although this application contains numerous jobs, each individual job is fine-grained in nature--in certain data scenarios it can be less than a minute. The overhead associated with initiating each task on a separate node and collecting its results after it finished execution would have radically decreased the efficiency of distributed execution. So, coarse grained jobs had to be created by grouping the fine-grained jobs so that the computation to I/O ratio would work in favour of distributed analysis. For this purpose, a meta-program (*metameg*) was written that would perform pairwise analysis for all sensors for a particular range of offsets. The entire temporal shift region was divided into offset ranges. A job would consist of performing analysis over a temporal shift range for all the sensors. The pseudo-code for this meta job program, called metameg, is listed in Fig. 4.

```
For time_offset from time_t1 to (time_t1+ time_offset_step)
begin
    For sensor_A from 1 to max_sensors
    begin
        For sensor_B from 1 to max_sensors
         begin
                if  sensor_A NOT EQUALS sensor_B then
                begin
                     Execute raw2wavelet sensor_A sensor_B time_offset meg_ data_path
                     Execute wavelet2cross sensor_A_output sensor_B_output
                end
         end
```

**Fig. 4: Pseudo-code for meta program**

The variable **time_offset_step** decides the size of the meta job as it divides the offset range into regions. If it is 1, then a job is the wavelet cross-correlation analysis for all sensors for one particular offset. If it is same as the limit of the temporal shift region (i.e. maximum offset), then it represents an aggregation of all the jobs. As the size of the meta job increases, the number of jobs generated decreases according to the following equation:

$$Meta\ \ Job\ Size = \frac{Maximum\ \ offset}{No\ of\ \ Meta\ \ Jobs\ \ for\ Grid\ \ Analysis}$$

The executables of the three programs that represented the analysis code (metameg, raw2wavelet, and wavelet2cross implemented in C language) were small in size and hence easily transferable at runtime if the code is not deployed on remote resources. Each meta job requires the full raw data captured by all (64) sensors. However, transferring this raw data, over 24MB, to each remote node during the execution any job would constitute a significant I/O load over wide-area networks. Unless an analysis is carried out on a MEG data with few sensors generating small amount of data, it is preferable to pre-stage the entire raw dataset at each node. We also had to change the meta program (metameg) and the raw data to wavelet conversion (raw2wavelet) program to enable access to data on a path on the remote node.

A Nimrod-G plan for the SPMD (single program and multiple-data) style execution of the above analysis operation as parametric execution is shown in Fig. 5. As the Grid comprises of heterogeneous resources with different architectures, we have compiled the analysis suite for different architectures and renamed the executables with an extension same as the OS name. That means, any reference to the program with .$OS in the plan file indicates that the Nimrod-G broker is able to select the correct executable depending on the target architecture automatically at runtime. $HOME represents the home directory of the user at the remote node and the directory $HOME/alphawave is where the brain data is stored on the remote machine.

The output of the each pair-wise analysis is represented in image data, which can be further studied using a visualization program jointly developed by the Osaka University and NEC, Japan. All the output files of meta job are grouped (archived using the tar command) and transferred back to the remote node.

```
parameter meg_sensors_count label "no. of MEG sensors" integer default 64;
parameter offset_max = 29750;
parameter MetaJobSize label "time_offset step size" integer default 10;
parameter time_offset label "MEG Data Temporal Region Shift" integer range from 0 to offset_max step MetaJobSize;
task nodestart
        # copy meg_data.tar node:$HOME/alphawave
        copy raw2wavelet.$OS node:raw2wavelet
        copy wavelet2cross.$OS node:wavelet2cross
        copy metameg-data-path.$OS node:metameg
endtask
task main
        node:execute ./metameg-data-path $time_offset $MetaJobSize $meg_sensors_count $HOME/alphawave
        node:execute $HOME/tar cvf output.tar *.asc *.ppm
        copy node:output.tar output.tar.$jobname
endtask
```

**Fig. 5: Plan file for brain activity analysis on the Grid.**

# 5. Scheduling Experiments and Results

The brain activity analysis was demonstrated at SC 2002 held in Baltimore, USA. We used the resources provided by the SC2002 Global Grid Testbed Collaboration [9], which we were part of, plus those of the World Wide Grid[7] to schedule analysis jobs on globally distributed resources. After the SC 2002 conference, most of the resources that were part of this testbed have discontinued their participation as the original objective has been achieved. This illustrates the ever changing nature of collaborations on the Grid. Different organisations are able to come together and share their resources to achieve an objective and thus create virtual organizations[10].

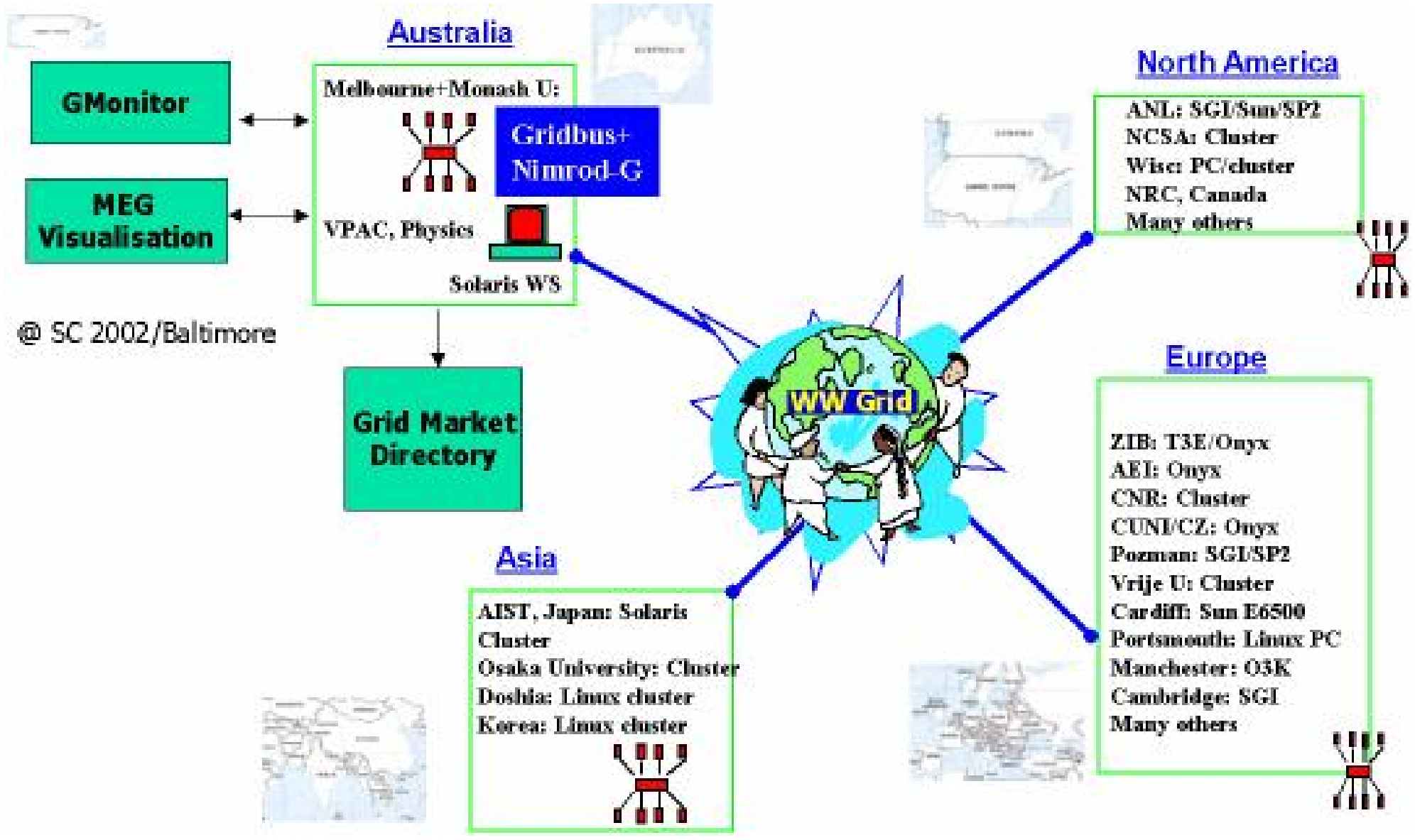

**Fig. 6: Brain Activity Analysis at the HPC Challenge Demo @ SC 2002.**

However, at the time of writing this paper (at the end of January 2003), we continued to have access to some of the testbed resources. We utilised these resources for our analysis experiments whose results are reported in this paper. The list of machines that we have used to carry out our experiments is shown in Table 1.

| **Organization** | **Node details (Architecture, No. of Nodes, Hostname)** | **Cost Per CPU-sec (G$)** |
|---|---|---|
| N*Grid Project Korea | `Linux Cluster, 24 nodes node1001.gridcenter.or.kr` | 3 |
| Vrije Universiteit, Netherlands | `Linux Cluster, 144 nodes(32 available),fs0.das2.cs.vu.nl` | 2 |
| N*Grid Project Korea | `Linux Cluster, 16 nodes, node2001.gridcenter.or.kr` | 1 |
| Osaka University, Japan | `Linux Cluster, 2 nodes, date1.ics.es.osaka-u.ac.jp` | 1 |
| Dept. of Physics, Uni. of Melbourne | `Linux, broker machine, lem.ph.unimelb.edu.au` | 0 |

**Table 1:** Grid resources utilised during the Brain activity analysis experiment.

The experiments were carried out on Thursday, $30^{th}$ of January 2003, from 10:00 a.m. to 1:00 p.m. AEDT. The three scheduling algorithms were tried out and their results tabulated. We performed the experiment for two sensors, maximum offset of 100 and meta-job size of 1. This would produce 100 jobs. With more jobs, there was a possibility that the state of the nodes would change during the conduct of the tests thereby causing inconsistent results. All experiments were started with: Deadline = 6hrs, Budget = 1990 Grid $. The summary of the results of these experiments is as follows:

| **Scheduling Strategy** | **Start Time** | **Completion Time** | **Budget Utilised (G$)** |
|---|---|---|---|
| Time | 10:00 a.m. | 10:29 a.m. | 399 |
| Cost | 10:35 a.m. | 11.54 a.m. | 204 |
| Cost-Time | 12:10 p.m. | 12.52 p.m. | 330 |

**Table 2:** Summary of experiment statistics.

The graph in Fig. 7 shows the progress of the execution using time minimization scheduling algorithm. As is depicted, the scheduler makes use of all the resources to ensure that the experiment completes in the fastest possible time. Most of the jobs were executed by the fastest machine (fs0.das2.vu.nl). However, it was also not the cheapest machine and so, the cost of computation increased.

The graph in Fig. 8 is of the same experiment using cost minimization scheduling. Here, most of the load is borne by the one of the cheapest machines (date.ics.es.osaka-u.jp). Since the deadline given is very long, the scheduler allocates all the jobs to the machine with the least cost to minimize the cost of computation. However, in the beginning, it does give some jobs to the faster, expensive machines before it can be sure that the Osaka University machine will be able to execute the jobs within the specified deadline.

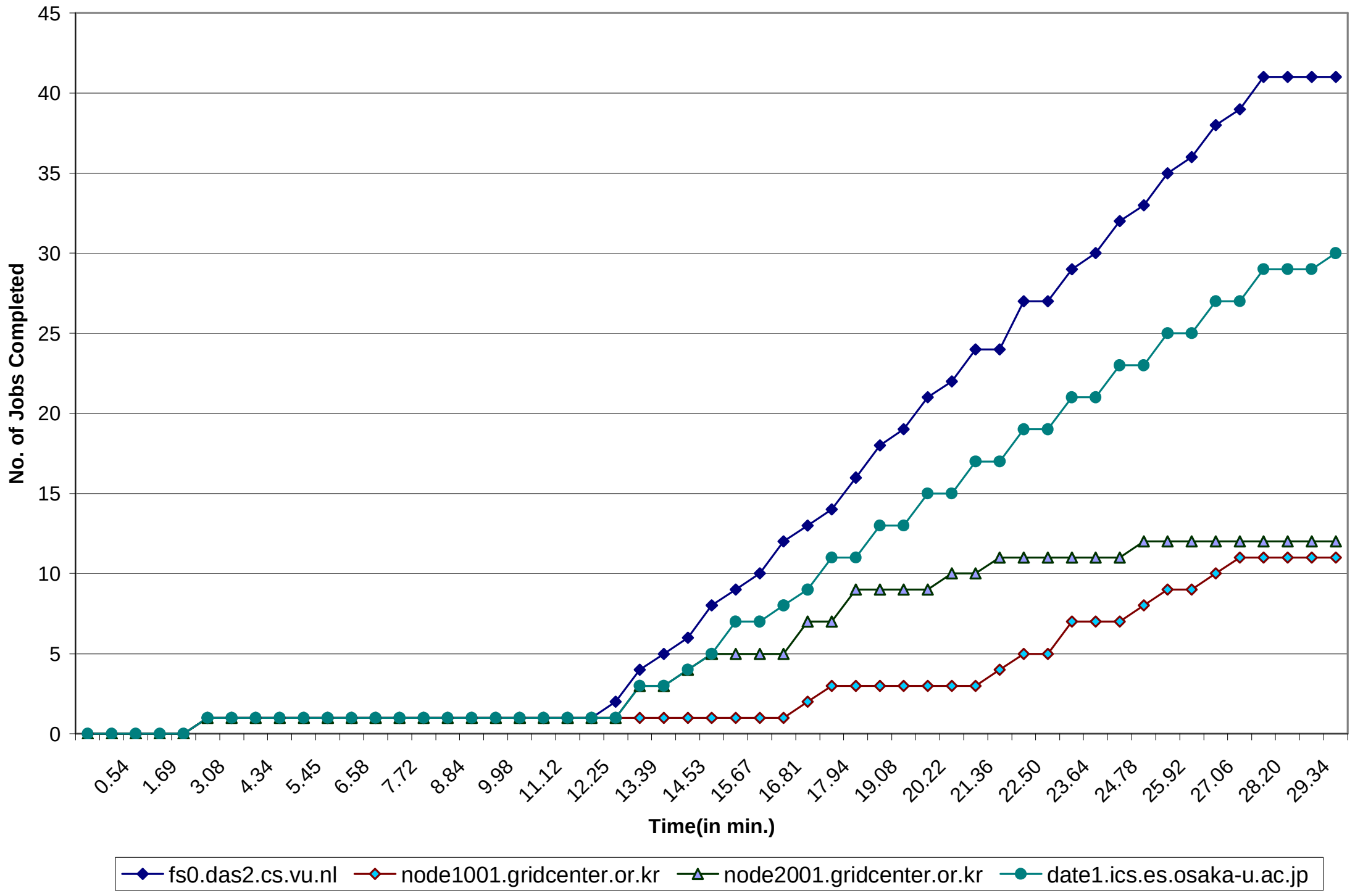

**Fig. 7: Scheduling with Time Minimization: Cumulative Graph of No. of Jobs Completed vs Time**.

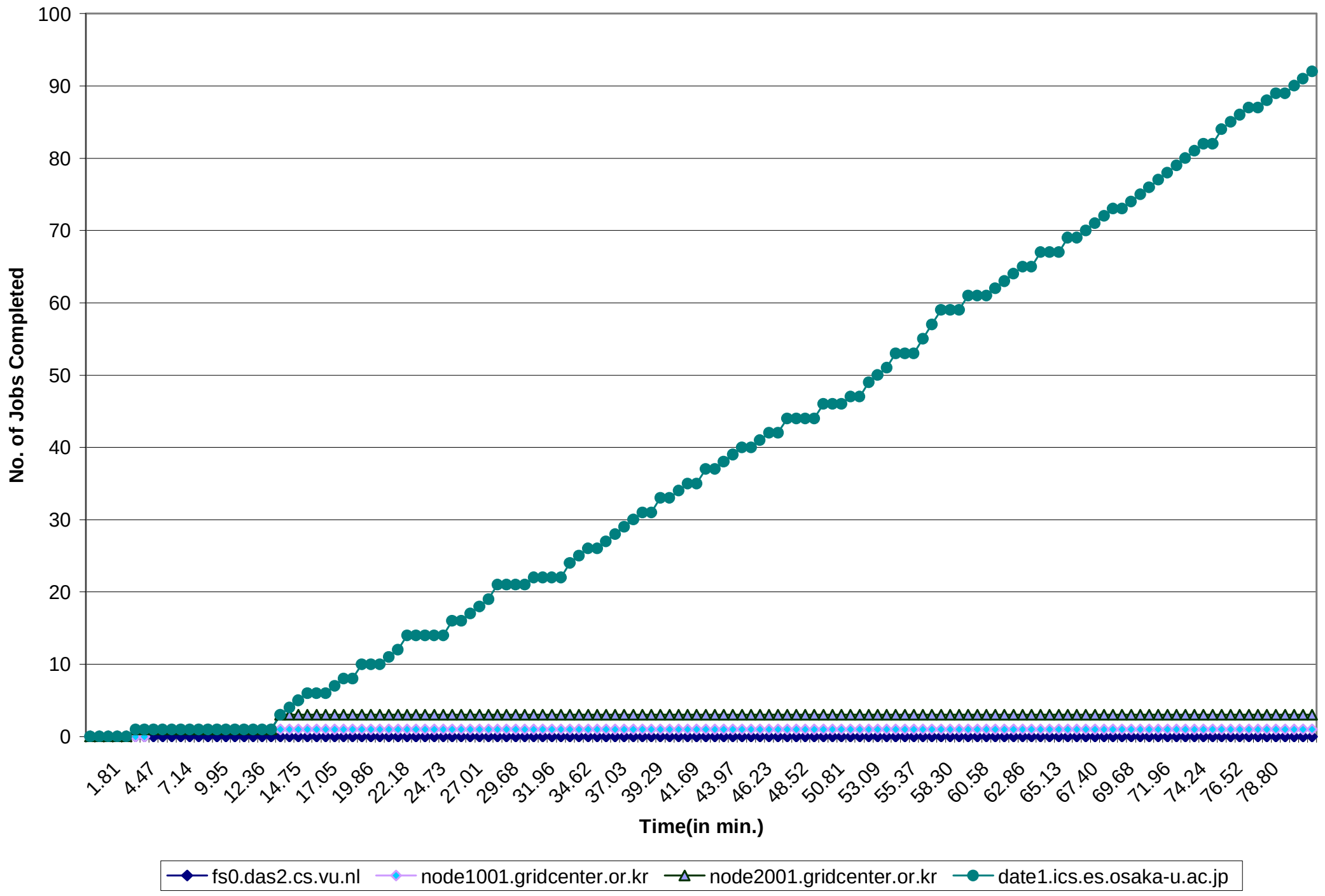

**Fig. 8 : Scheduling with Cost Optimization: Cumulative Graph of No. of Jobs Completed vs Time.**

The graph in Fig. 9 relates to the experiment performed with cost-time optimization. Here, most of the computation was handled by the cheapest nodes (node1001.gridcenter.or.kr and date1.ics.es.osaka-u.jp). But, initially as these two were busy, the scheduler allocated the maximum number of jobs to the expensive machines. As the availability of the other two improved, the share of jobs given to the costly machines decreased as jobs were moved to the cheaper resources. The graph does not show the behaviour expected of the algorithm that was observed in simulation[7]. This underscores the volatility of the Grid environment due to the constantly shifting loads on the machines.

The results of our scheduling experiments show that it is feasible to conduct the analysis using the parameter sweep model of distributed computing. Furthermore, this approach allows medical personnel to determine the pace of the analysis based on the urgency of their requirements. If the urgency is low, for example, the analysis might be conducted as a part of a study or for a report, then it is possible to reduce expenditure by using cost-optimization and a relaxed deadline. However, if the results are required in the shortest possible time, then the analysis can be performed using time optimization algorithm.

The visualization of wavelet analysis results of selected sensors is shown in Fig. 10 using the wavelet viewer. The viewer also has the ability to invoke wavelet analysis module on the local machine if the results corresponding to the selected sensor are not available. Wavelet analysis decomposes the frequency components of original data over the time. This is the reason why this analysis is promising in comparison with other frequency analysis such as Fourier. Most of traditional analysis methods for the analysis of brain function loses the time information of the original data. Therefore, medical doctors had much difficulty in investigating the change in frequency of MEG data over the time. The visualization was performed so that maximum value of the results is visualized as red, while minimum value of the results is visualized as blue. We expect that medical doctors can get the intuitive understanding of results at a glance.

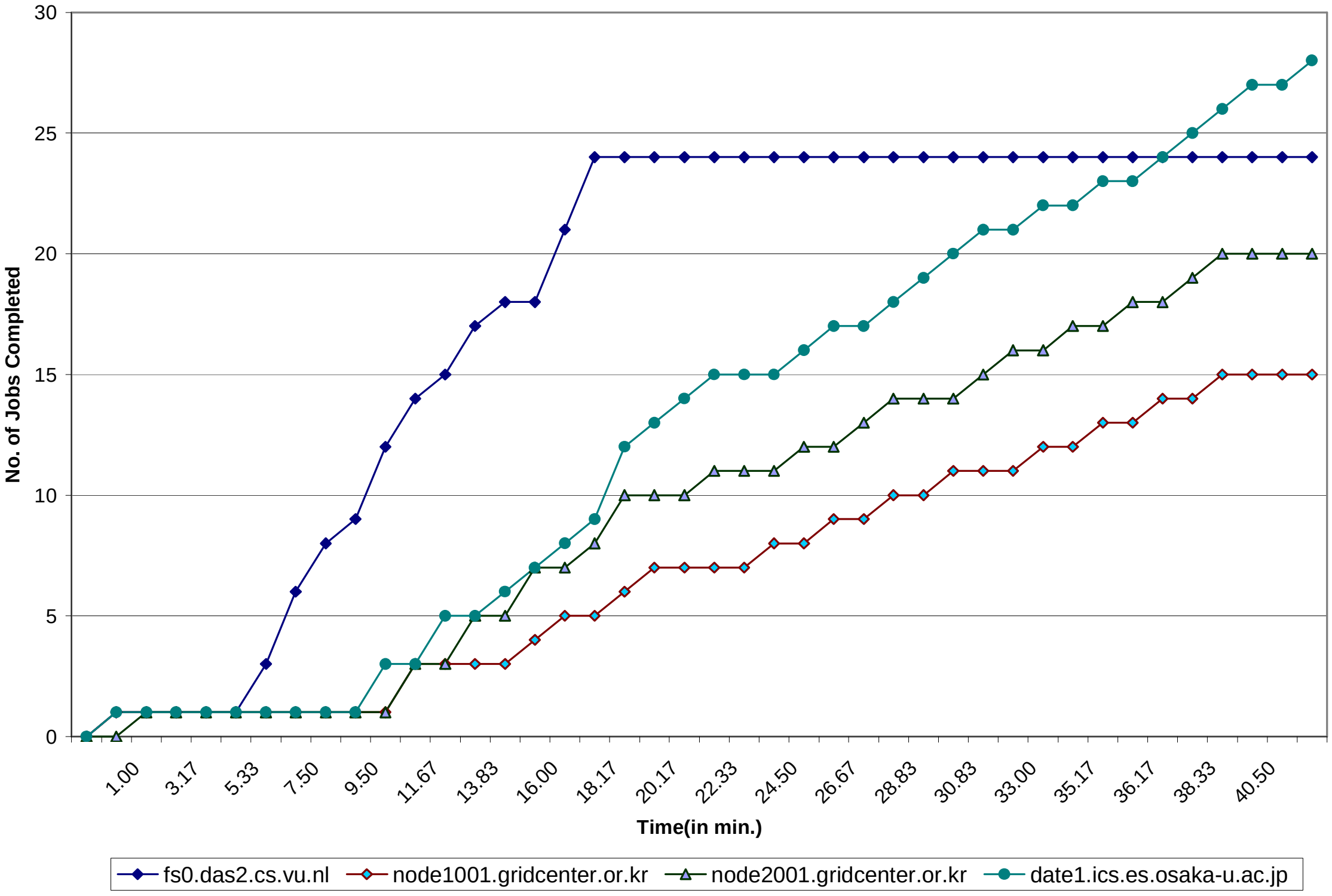

**Fig. 9: Scheduling with Cost-Time Optimization: Cumulative Graph of No. of Jobs Completed vs Time.**

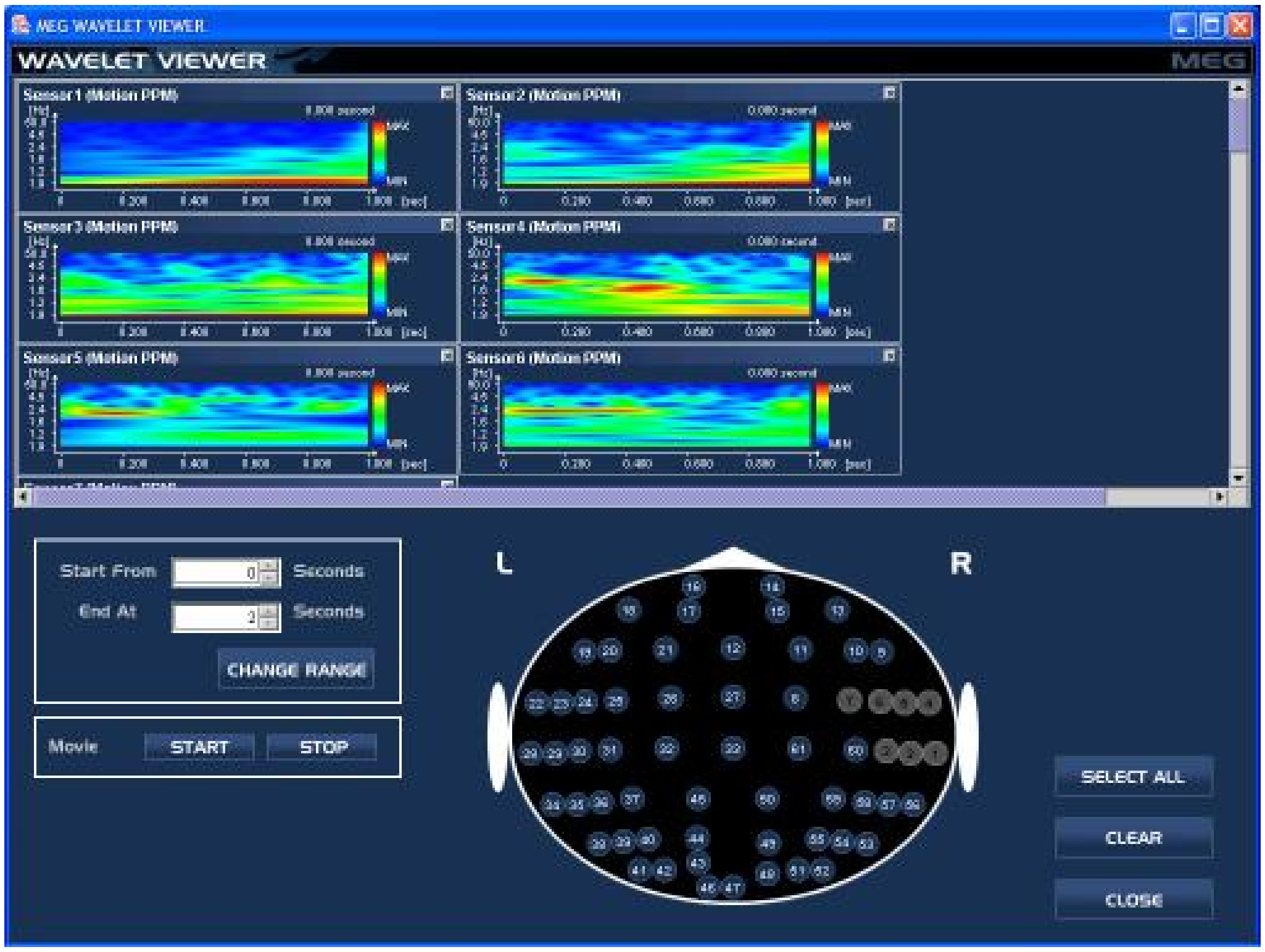

**Fig. 10: Visualisation of wavelet analysis results for selected sensors. The numbers within oval shape diagram indicate the sensor number.**

## 6. Related Work

The Osaka University has implemented a grid enabled version of the brain activity analysis application[2, 11] using MPICH-G and Globus. It has been observed that developing an MPI version of the program took significantly more resources (more than 6 months) compared to our solution. Hence, it can be safely said that the deployment time of the application was reduced considerably by adopting the task farming approach for distributed execution of brain activity analysis application. Additionally, the approach presented in this paper offered QoS based deployment of analysis jobs on Global Grids.

The Grid Research and Innovation Laboratory at University of California, San Diego has done application level scheduling for parallel tomography application[12]. However, the scheduler that they created was application-specific but our project aims to use a generic scheduler for executing the analysis jobs. Other related works include FightAIDS@Home project[13], which is based on the Entropia's distributed computing network and the Scripps Research Institute's docking application. In this system, volunteers need to download Entropia's screen saver program that runs in the background on the volunteer computer. The volunteer PC contacts the Entropia server to download the data to perform docking. When docking on an assigned data is completed, it uploads the results to the server. This execution model is different from our model where the scheduler (Nimrod-G) assigns the work to computers that are available and initiates the execution.

## 7. Conclusion and Future Work

As the new generation of medical instruments turns out to be more precise and accurate, the medical field will sooner or later have to find the means to deal with large volumes of data generated by them. Grid computing can satisfy their needs for a large amount of processing power but would have to deal with issues of tight deadlines, small turnaround time, consistency in performance and reliability of computation before any large scale adoption. The economy based approach of processing brain activity data as illustrated in this paper would help in enforcing QoS requirements of medical applications and hence would enable adoption of Grid technologies by the bio-instrumentation field.

We are in the process of writing a new job dispatcher which would allow us to group similar jobs and dispatch the group as a single job to a remote node. This would allow us to gather data from the instrument as and when available and dispatch it for processing on the Grid immediately thus creating truly on-demand brain activity analysis. We also plan to integrate our scheduler with accounting mechanisms such as GridBank and more tightly with the Grid Market Directory to take complete advantage of their capabilities. We also plan to support economy based advance reservation of Grid resources.

## Acknowledgement

We would like to thank participants of the SC2002 Global Grid Testbed Collaboration for providing us access to the resources that have made large scale experiments possible. We are grateful to all the system administrators of the machines that were part of the collaboration who have responded to our requests and have made it possible for our applications to execute on their systems. We thank Ed Seidel of AEI-Potsdam, Germany for inviting us to join this testbed collaboration. We would also like to thank Slavisa Garic of Distributed Systems Technology Centre, Monash University, Australia for having made changes to Nimrod-G so that we could use as many resources as possible of those that were available at our disposal. Also, this study was partly supported by IT-program of Ministry of Education, Culture, Sports, Science and Technology (MEXT), Japan. And we would like to thank Life-Electronics laboratory, National Institute of Advanced Industrial Science and Technology (AIST) for the use of the MEG.